\documentclass[twocolumn,showpacs,preprintnumbers,amsmath,amssymb]{revtex4}

\usepackage{graphicx}
\usepackage{dcolumn}
\usepackage{bm}

\input{pstricks.tex}
\input{pst-plot.tex}
\input{pst-coil.tex}
\input{pst-node.tex}

\begin{document}


\title{Exact  calculation of  the  energy contributions  to the  $T=0$
random-field  Ising  model  with  metastable  dynamics  on  the  Bethe
lattice}

\author{Xavier   Illa}   
\email{xit@ecm.ub.es}  
\author{Jordi   Ort\'{\i}n}
\email{ortin@ecm.ub.es} 
\author{Eduard Vives} 
\email{eduard@ecm.ub.es}
\affiliation{ Departament d'Estructura  i Constituents  de  la Mat\`eria,
Facultat  de F\'{\i}sica, Universitat  de Barcelona  \\ Diagonal  647, 
08028 Barcelona, Catalonia (Spain)}

\date{\today}

\begin{abstract}
We analyze  the energy terms  corresponding to the  spin-spin exchange
$\sum S_i  S_j$ and spin-random field  coupling $\sum S_i  h_i$ of the
zero temperature random-field Ising  model on the Bethe lattice driven
by  an external  field with  metastable dynamics.   Exact  results are
calculated as  a function  of the standard  deviation of  the disorder
$\sigma$ and the coordination  number $z$, and compared with numerical
simulations on  random graphs for $z=4$, for  which a disorder-induced
transition takes place.
\end{abstract}



\maketitle

\section{Introduction}
\label{Intro}
Hysteresis   and   metastability   are   intriguing   phenomena   with
implications    in    both    fundamental    and    applied    physics
\cite{Bertotti1998}. They  arise as a consequence of  the existence of
internal  energy barriers  that cannot  be overcome  by the  system. A
particularly  interesting case is  the so  called ``rate-independent''
hysteresis, for which  metastability is not a consequence  of the fast
driving  rate but  of the  ``athermal'' character  of the  system. The
energy barriers  are so high  compared with thermal  fluctuations that
effectively  the  system  behaves  at zero  temperature,  following  a
reproducible  and deterministic metastastable  path when  the external
field is  varied. Many  experimental situations are  known to  be well
approximated by this extreme case.

Several  models   have  been  useful  for   the  characterization  and
description of  rate independent hysteresis  in different experimental
systems.   A very interesting  microscopic model  is the  Random Field
Ising   model  at   $T=0$  with   metastable  dynamics   ($T=0$  RFIM)
\cite{Sethna1993}.    It  contains   the  three   essential  competing
ingredients   for  the   occurence  of   ``athermal''   hysteresis:  a
ferromagnetic nearest-neighbour (n.n.) interaction term favouring long
range order,  a local energy  term associated with  quenched disorder,
and a term describing the coupling of the magnetization to an external
driving  field.  Although  the model  is formulated  using  a magnetic
language  it can  be  translated easily  to  other systems  displaying
athermal hysteresis.

During  the last  decade  the $T=0$  RFIM  model has  been  used as  a
prototype  for  the understanding  of  many  properties associated  to
hysteresis:   return  point   memory,   congruency,  distribution   of
metastable  states  \cite{Basso2004,Detcheverry2004},  demagnetization
process   \cite{Dante2002,Colaiori2002},  disorder   induced  critical
points    \cite{Dahmen1993,Dahmen1996,Perkovic1999}    and   power-law
distribution  of  the   magnetization  avalanches  (Barkhausen  noise)
\cite{PerezReche2003,PerezReche2004b}.  Most of these results are based
on numerical simulations on finite lattices or on mean field analysis.
Interestingly,  however,  non  trivial  analytical  solutions  can  be
obtained for the  particular case of the $T=0$  RFIM on Bethe lattices
with  coordination number $z$.   The main  hysteresis loop  was solved
seven years ago by Shukla, Dhar and Sethna \cite{Shukla1996,Dhar1997}.
More   recently   partial   loops   \cite{Shukla2000,Shukla2001}   and
trajectories   starting   from    states   with   ``quenched''   spins
\cite{Shukla2004}  have  also been  deduced  .   Here  we present  the
explicit computation  of the three energetic terms  of the Hamiltonian
${\cal H}$ for a generic values of $\sigma$ and $z$
\footnote{During the  final stages of this  work, we were  aware of an
unpublished  document  by L.Dante  that  contains  basically the  same
calculation restricted to the case $z=2$.}. Analytical results for all
the contributions to ${\cal H}$ give insight on the singular behaviour
of the system at $\sigma_c$,  and allows for a deeper understanding of
the energy balances in the hysteresis loop.

The paper is organized as follows. In section \ref{Model} we summarize
the  details of  the model.   In  section \ref{General}  we solve  the
different terms in the Hamiltonian. In section \ref{z4} we present the
results for the  $z=4$ case and compare with  numerical simulations on
random graphs.  Finally, in section \ref{dissip}, we  study the energy
dissipation.

\section{Model}
\label{Model}
The RFIM is defined on a Bethe lattice with $N$ sites and coordination
number  $z$.  On each  lattice site  we define  a spin  variable $S_i$
which  takes  values $\pm  1$.   The  Hamiltonian (magnetic  enthalpy)
reads:
\begin{equation}
{\cal H}  = +U_{e}+U_{d}  - HM
\label{hamiltonian}
\end{equation}
where $H$ is the external driving field,
\begin{equation}
M(\{S_i\})= \sum_{i=1}^{N} S_i
\end{equation}
is the magnetitzation,
\begin{equation}
U_{e}(\{S_i\}) = -J\sum_{\langle i,j \rangle} S_i S_j 
\end{equation}
is the ferromagnetic exchange energy extending over all n.n. pairs, and
\begin{equation}
U_{d}(\{S_i\}) = -\sum_{i=1}^{N} h_i S_i
\end{equation}
accounts for the energy interaction with quenched disorder.
The random fields $\{h_i\}$ are independent and distributed according to a
Gaussian probability density centered around zero:
\begin{equation}
f(h_i) = \frac{1}{\sqrt{2 \pi} \sigma} e^{-\frac{h_i^2}{2 \sigma^2}}
\end{equation}
where  $\sigma$ is  the standard  deviation of  the random  fields and
controls the  amount of  disorder in the  system.  

For  the analysis  of  metastability  and hysteresis  loops  we use  a
1-spin-flip local  relaxation dynamics.   This is the  standard choice
used   in   previous   studies    of   the   metastable   $T=0$   RFIM
\cite{Sethna1993}: each spin $S_i$ flips individually according to the
sign of its local field $F_i$ given by:
\begin{equation}
F_i = J\sum_{j=1}^z S_j + H + h_i
\label{localfield}
\end{equation}
where the  first sum  extends over the  $z$ neighbours of  $S_i$.  The
complete  lower   branch  of  the  hysteresis  loop   is  obtained  by
adiabatically  increasing $H$  from $-\infty$  $(M=-N)$  to $+\infty$
$(M=+N)$.

In order  to check  the analytical results  that will be  presented we
have  also  performed  numerical  simulations on  random  graphs  with
coordination  number $z=4$  and  a  range of  sizes  from $N=10^4$  to
$N=10^6$ .  It is known that in the thermodynamic limit such numerical
simulations  agree  with the  analytical  results  for Bethe  lattices
\cite{Dhar1997}.  For  the simulations  we start with  a value  of $H$
negative enough  so that the unique  stable state is given  by all the
spins  $S_i=-1$.   We increase  the  external  field  $H$ until  $F_i$
vanishes  on a certain  spin. The  spin is  then reversed  keeping $H$
constant.   This reversal  may  destabilize some  of the  neighbouring
spins,  which are  then reversed  simultaneously  (parallel updating).
This is the begining of  an avalanche.  The avalanches proceed until a
new metastable situation  with all the spins $S_i$  aligned with their
respective  local  fields $F_i$  is  reached.   We  can then  continue
increasing the external field $H$.

In the figures presented below numerical simulations correspond always
to  averages over  several ($\sim  10$)   realizations of  the random
graph and many realizations of the random fields ($\sim 1000$).

\section{General solution for coordination number $z$}
\label{General}
Our goal  is the computation of  the different energetic  terms in the
Hamiltonian.  The  average over realizations  of the random  fields of
the three terms in the Hamiltonian gives:
\begin{equation}
\frac{\left< -HM \right>}{N} =-H\left< S_i \right> \equiv -Hm
\end{equation}
\begin{equation}
\frac{\left<  U_{e}  \right>}{N}  =-\frac{1}{2}  zJ \left<  S_i  S_j\right>
\end{equation}
\begin{equation}
\frac{\left< U_{d} \right>}{N} = -\left< h_i S_i \right>
\end{equation}
The averages on the right hand sides can be computed as follows:
\begin{equation}
\left< S_i \right>=\sum_{\{S_i\}} S_i P(S_i)=P(+1)-P(-1)
\label{si}
\end{equation}
where $P(S_i)$ is the probability for a spin to take a value $\pm 1$,
\begin{eqnarray}
\left< S_i S_j \right>&=&\sum_{\{S_i,S_j\}} S_i S_j P(S_i,S_j)=   \nonumber \\
&=&P(+1,+1)+P(-1,-1)-2P(+1,-1)     \nonumber\\
& &                             
\label{sisj}
\end{eqnarray}
where $P(S_i,S_j)$ is the probability  for a nearest neighbour pair to
be   in   the    state   $(S_i,S_j)$.   This   probability   satisfies
$P(+1,-1)=P(-1,+1)$. Finally
\begin{eqnarray}
\left< h_i S_i \right>=\sum_{\{S_i\}} \int_{-\infty}^{+\infty} dh_i P(h_i,S_i) h_i S_i= \nonumber \\
\int_{-\infty}^{+\infty} dh_i P(h_i,+1) h_i -   \int_{-\infty}^{+\infty} dh_i P(h_i,-1) h_i
\label{hisi}
\end{eqnarray}
where $P(h_i,S_i)$ is the probability  for a site $i$ of having
a random field within $(h_i,h_i+dh_i)$ and a spin with state $S_i$.

\begin{figure}[ht]
\begin{center}
\begin{pspicture} (8,10)(1,1.5)
\psset{unit=0.7 cm}
\put(1,12.5){(a)}
\psdots[dotstyle=*,dotscale=2.0](4,11)
\psdots[dotstyle=*,dotscale=2.0](6,11)
\psdots[dotstyle=*,dotscale=2.0](8,13)(8,12)(8,9)
\put(5.8,10.3){$S_i$}
\psline(6,11)(4,11)
\psline(6,11)(8,13)
\psline(6,11)(8,12)
\psline(6,11)(8,9)
\psdots[dotstyle=*,dotscale=0.5](8,11)(8,10.5)(8,10)
\put(8.5,10.5){$z-1$} 
\put(4.6,9){$z$ neighbours}
\put(4.6,8.5){$n$ spins $+1$}
\put(1,7.5){(b)}
\psdots[dotstyle=*,dotscale=2.0](3,7)(3,6)(3,3)
\psdots[dotstyle=*,dotscale=2.0](5,5)(7,5)
\put(4.8,4.4){$S_i$}
\put(6.8,4.4){$S_j$}
\psdots[dotstyle=*,dotscale=2.0](9,7)(9,6)(9,3)
\psline(5,5)(7,5)
\psline(5,5)(3,7)
\psline(5,5)(3,6)
\psline(5,5)(3,3)
\psline(7,5)(9,7)
\psline(7,5)(9,6)
\psline(7,5)(9,3)
\psdots[dotstyle=*,dotscale=0.5](3,5)(3,4.5)(3,4)
\psdots[dotstyle=*,dotscale=0.5](9,5)(9,4.5)(9,4)
\put(1,4.5){$z-1$}
\put(9.5,4.5){$z-1$}
\put(2,2){$l$ spins $+1$}
\put(8,2){$r$ spins $+1$}
\end{pspicture}
\caption{\label{fig1}  Schematic representation  of the  enviroment of
(a) a single site and (b) a pair.  The variables $n$, $l$, $r$ account
for the number of spins $+1$. }
\end{center}
\end{figure}
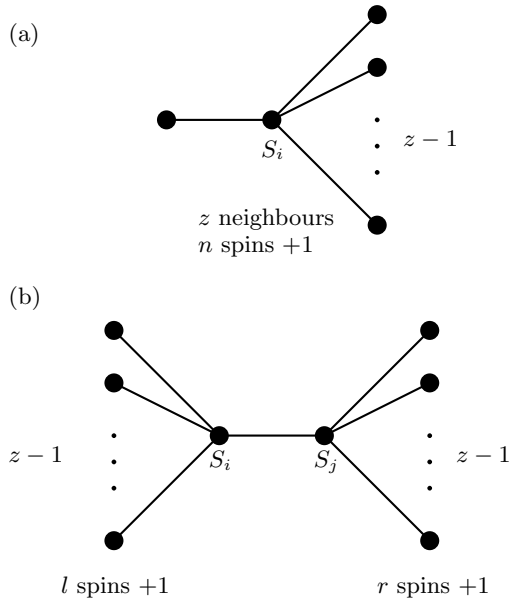

At  this point  we follow  Ref.~\onlinecite{Dhar1997}.  We  define the
probability  $P(S_i |n)$  for  a spin  being  in state  $S_i$ given  a
certain environment of nearest  neighbours.  This environment is fully
characterized by the variable $n$ ($0 \le n \le z$) which accounts for
the     number     of     neighbours     in    state     $+1$     (see
Fig.~\ref{fig1}(a)). Clearly,
\begin{eqnarray*}
P(S_i=+1 |n)=\int_{-J(2n-z)-H}^{+\infty} dh_i f(h_i)= \\
=\frac{1}{2} \mbox{erfc}\left\{\frac{-J(2n-z)-H}{\sqrt{2}\sigma^2 }\right\} 
\end{eqnarray*}
and
\begin{eqnarray*}
P(S_i=-1 |n)=1-P(S_i=+1 |n)
\end{eqnarray*}
From Bayes formula one can write:
\begin{equation}
P(S_i)=\sum_{n=1}^{z} P(n) P(S_i | n)
\label{ps}
\end{equation}
where $P(n)$ is  the probability for a site  having an enviroment with
$n$  neighbouring spins  in the  state $+1$  (see Fig.~\ref{fig1}(a)).
According to Ref.~\onlinecite{Dhar1997},
\begin{equation}
P(n)=\left (\begin{array}{c} z \\ n\end{array} \right)
{P^{\ast}}^{n}(1-P^{\ast})^{(z-{n})}
\label{penv}
\end{equation}
where  $P^{\ast}$ results  from defining  the  conditional probability
that  a spin  is $+1$  given that  its parent  spin (according  to the
hierarchy of the Bethe lattice)  is down, and its descendent spins are
relaxed.   In   a  site  deep   enough  inside  a  very   big  lattice
(thermodynamic  limit), this  probability tends  to be  homogeneus and
$P^{\ast}$ is given by the self consistent equation \cite{Dhar1997}:
\begin{equation}
P^{\ast}=\sum_{n=0}^{z-1} P^{\ast n} (1-P^{\ast})^{z-1-n} P(S_i=+1 |n) 
\label{past}
\end{equation}
Note that  this equation implicitely  contains the information  on the
fact that we  are increasing the field monotonously  from the negative
fully saturated state.

By  numerically   obtaining  $P^\ast$  from   (\ref{past})  and  using
(\ref{penv}) and (\ref{ps}),  the averaged magnetization in (\ref{si})
can be obtained.   This allows to calculate the  averaged lower branch
of the hysteresis loop.

In order to  compute the terms (\ref{sisj}) and  (\ref{hisi}) we apply
Bayes formula again and write
\begin{equation}
P(S_i,S_j)=\sum_{l=1}^{z-1}\sum_{r=1}^{z-1}  P(l,r) P(S_i,S_j | l,r)
\label{psisj}
\end{equation}
\begin{equation}
P(h_i,S_i)=\sum_{n=1}^{z} P(n) P(h_i,S_i | n) \;,
\label{phisi}
\end{equation}
where  $P(S_i,S_j   |  l,r)$  and  $P(h_i,S_i  |   n)$  are  conditional
probabilities  given  a  certain   environment  and  $P(l,r)$  is  the
probability for  a pair  having an environment  with $l$ spins  in the
state $+1$ in  the left neighbourhood and $r$ spins  in the state $+1$
in  the  right  neighbourhood  (see  Fig.~\ref{fig1}(b)).   This  is  the
generalitation of $P(n)$  for the description of the  environment of a
pair of spins.

\subsection{Calculation of $P(S_i,S_j)$}
The calculation of $P(S_i,S_j)$ starts by generalizing Eq.~(\ref{penv}) to  
\begin{eqnarray}
P(l,r)& = &g(l,r) {P^{\ast}}^{l}(1-P^{\ast})^{(z-1-l)}{P^{\ast}}^{r}(1-P^{\ast})^{(z-1-r)}= \nonumber \\
&=&g(l,r) {P^{\ast}}^{l+r}(1-P^{\ast})^{(2z-2-r-l)}
\label{penv2}
\end{eqnarray}
where
\begin{equation}
g(l,r)=\left (\begin{array}{c} z-1 \\ l
\end{array} \right) \left (\begin{array}{c} z-1 \\ r
\end{array} \right)
\end{equation}

The  conditional probabilites  for a  pair  of spins  given a  certain
environment, $P(S_i,S_j | l,r)$, can be written as double integrals of
the random field  distribution on a certain domain in  the $h_i - h_j$
plane, i.e.
\begin{equation}
P(S_i, S_j | l,r)=\int \int_{D} f(h_i)f(h_j)dh_i dh_j
\label{psisjlr}
\end{equation}
Figure \ref{plahihj} shows an example of such domains corresponding to the
environment  $l=1$   and  $r=1$.  The  plots   corresponding  to  other
environments are obtained by translation of this case. Therefore:
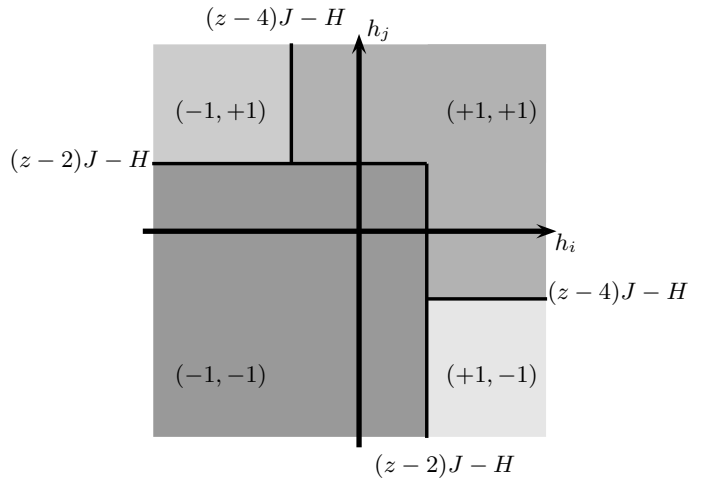
\begin{figure}
\begin{center}
\begin{pspicture} (5,5) (-0.5,-0.5) 
\psset{unit=0.5 cm}
\newgray{gray1}{0.60}
\newgray{gray2}{0.70}
\newgray{gray3}{0.80}
\newgray{gray4}{0.90}
\psframe[linestyle=none,fillstyle=solid,fillcolor=gray1](-0.5,-0.5)(6.8,6.8)
\psframe[linestyle=none,fillstyle=solid,fillcolor=gray2](3.2,6.8)(10,10)
\psframe[linestyle=none,fillstyle=solid,fillcolor=gray2](6.8,3.2)(10,10)
\psframe[linestyle=none,fillstyle=solid,fillcolor=gray3](-0.5,6.8)(3.2,10)
\psframe[linestyle=none,fillstyle=solid,fillcolor=gray4](6.8,-0.5)(10,3.2)
\psline[linewidth=1.3pt](3.2,6.8)(3.2,10)
\psline[linewidth=1.3pt](6.8,-0.5)(6.8,6.8)
\psline[linewidth=1.3pt](6.8,3.2)(10,3.2)
\psline[linewidth=1.3pt](-0.5,6.8)(6.8,6.8)
\psline[linewidth=2.0pt]{->}(5,-0.75)(5,10.25)
\psline[linewidth=2.0pt]{->}(-0.75,5)(10.25,5)
\put(10.2,4.5){$h_i$}
\put(5.2,10.2){$h_j$}
\put(0.9,10.5){$(z-4)J-H$}
\put(5.4,-1.4){$(z-2)J-H$}
\put(10.0,3.2){$(z-4)J-H$}
\put(-4.3,6.7){$(z-2)J-H$}
\put(7.3,8.0){$(+1,+1)$}
\put(7.3,1.0){$(+1,-1)$}
\put(0.1,8.0){$(-1,+1)$}
\put(0.1,1.0){$(-1,-1)$}
\end{pspicture}
\caption{\label{plahihj}  The $(h_i,h_j)$  plane divided  in different
areas  (different color)  corresponding to  the final  state  of spins
$(S_i,S_j)$ for  a fixed enviroment $l=1,r=1$.  These areas correspond
to the domains D of integration in Eq.~(\ref{psisjlr}).}
\end{center}
\end{figure}
\begin{equation}
P(-1, -1 |  l,r)=\left [ 1-P(S_i=+1 |l) \right ] \left [1- P(S_i=+1 |r)\right ]
\end{equation}
\begin{equation}
P(+1, -1 | l,r)= P(S_i=+1 |l)  \left [1- P(S_i=+1 |r+1)\right ]
\end{equation}
\begin{equation}
P(-1, +1 | l,r)=\left [ 1-P(S_i=+1 |l+1) \right ] P(S_i=+1 |r)
\end{equation}
\begin{eqnarray}
P(+1, +1 | l,r)=P(S_i=+1|l+1)P(S_i=+1|r)+ \nonumber \\
\; + P(S_i=+1|l) \left[ P(S_i=+1 |r+1)-P(S_i=+1 |r) \right]
\end{eqnarray}
From these equations and using (\ref{penv2}) and (\ref{psisj}) one can
obtain  the  correlation $\langle  S_i  S_j  \rangle$  needed for  the
computation of Eq.~(\ref{sisj}).
 
\subsection{Calculacion of $P(h_i,S_i)$}
The  computation  of   $P(h_i,S_i)$  is  straightforward  noting  that
$P(h_i,S_i| n)$ can be computed  as tails of the gaussian distribution
$f(h_i)$ as:
\begin{equation}
P(h_i,+1 | n)= \left \{ \begin{array}{lcr} 0     &   & h_i < (z-2n)J-H \\
                                                          f(h_i)  &   & h_i > (z-2n)J-H  \end{array} \right.
\end {equation}
\begin{equation}
P(h_i,-1 | n)= \left \{ \begin{array}{lcr} f(h_i) &   & h_i < (z-2n)J-H \\
                                                              0    &   & h_i > (z-2n)J-H  
                                         \end{array} \right.
\end {equation}
Thus, using  (\ref{penv}) and  (\ref{phisi}) one can  compute $\langle
h_i S_i \rangle$ needed for the computation of Eq.~(\ref{hisi}).

\section{Numerical solution for the case $z=4$}
\label{z4}
The case with  coordination number $z=4$ is interesting  because it is
known \cite{Dhar1997}  to display a disorder  induced phase transition
between smooth  hysteresis loops  for $\sigma >  \sigma_c=1.78125$ and
discontinuous hysteresis  loops for  $\sigma < \sigma_c$.   We present
the  results  obtained by  numerically  solving  the  equation of  the
previous  section  for  $z=4$.    In  particular  the  real  roots  of
Eq.~(\ref{past}) are  found with a bisection method  restricted to the
interval $0\leq P^{\ast}\leq 1$.

\begin{center}
\begin{figure}[th]
\includegraphics[width=8.5cm,clip]{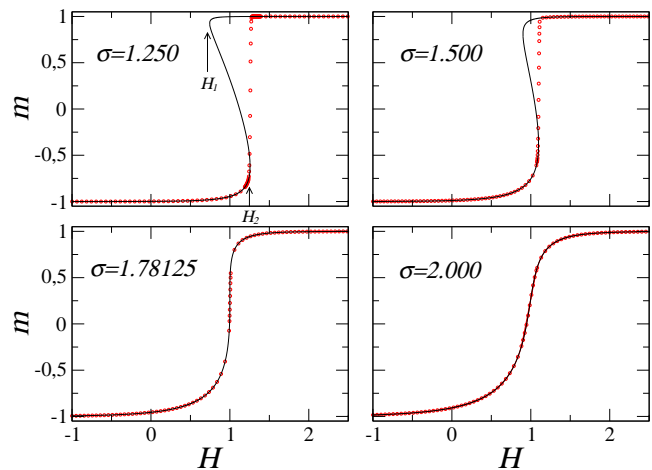}
\caption{\label{fig3}   Lower   branches   of  the   hysteresis   loop
corresponding  to  $z=4$ and  different  amounts  of disorder.   Exact
results  (continuous lines)  are compared  with  numerical simulations
(dotted lines). Simulations correspond to  a random graph with $N=10^5$
and averages over $1000$ different realizations of disorder }
\end{figure}
\end{center}

Figure  \ref{fig3} shows  the  $m$-$H$ diagram  corresponding to  four
different  amounts   of  disorder   $\sigma$.   Note  that   the  data
represented  corresponds only to  the lower  branch of  the hysteresis
loop for increasing the external field $H$.

\begin{center}
\begin{figure}[th]
\includegraphics[width=8.0cm,clip]{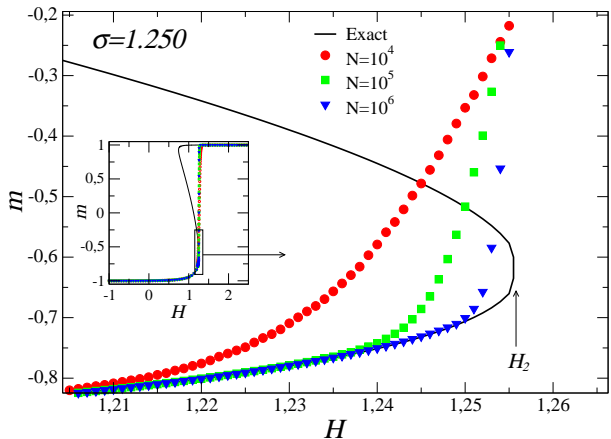}
\caption{\label{fig4} Examples of finite  size dependence of the lower
branch of  the hysteresis loop.  Symbols correspond  to simulations on
random graphs  with increasing sizes  as indicated by the  legend. The
continuous line corresponds to the exact solution.}
\end{figure}
\end{center}

The  disconinuity  in  the  $m(H)$  branch,  as  was  pointed  out  in
Ref.~\cite{Dhar1997},  arises  from  the  fact that  the  solution  of
equation (\ref{past}) is trivalued  for $\sigma<\sigma_c$ in a certain
field range $H_1<H<H_2$.  The three roots of Eq.~(\ref{past}) generate
the  s-shape  curve in  the  $m-H$ diagram  than  can  be observed  in
Fig.~\ref{fig3}   for   $\sigma<\sigma_c$.   Nevertheless,   numerical
simulations show that  only one of the roots  has physical meaning. In
the case of increasing field, only the lower $m$ branch is obtained in
the  simulations, and  the discontinuity  occurs at  $H_2$  where this
lower branch of  the s-shape curve joints the  intermediate branch and
disappears.  To  our knowledge there is no  clear physical explanation
for this fact and, a priori, the  jump could occur at any field in the
range $H_1$-$H_2$.  A stability criterium would be desirable.  In this
respect we note that in a very recent paper \cite{Detcheverry2004}, it
has been speculated that the  s-shape curve is related to the boundary
of the  density of 1-spin-flip metastable  states \cite{Vives2004}. It
is also important  to notice that the result  of numerical simulations
depends  on system  size, as  shown in  Fig.~\ref{fig4}.  Only  in the
thermodynamic limit  $(N \rightarrow \infty)$ the  data from numerical
simulations would  follow exactly the lower branch  of the theoretical
curve up to $H_2$.

\begin{center}
\begin{figure}[th]
\includegraphics[width=8.5cm,clip]{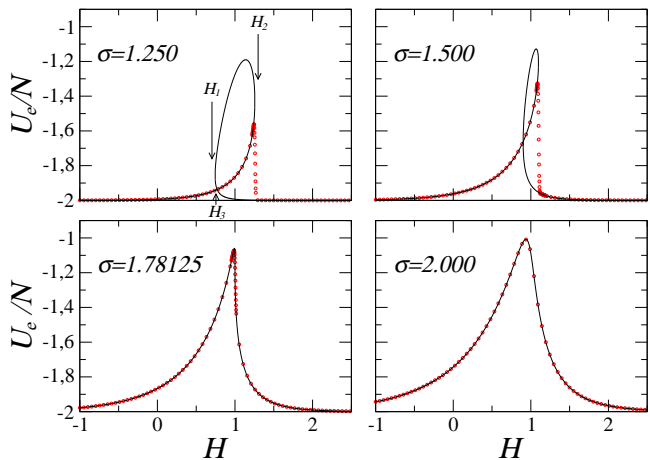}
\caption{\label{fig5} Exchange  energy behaviour corresponding  to the
same  cases  as   in  Fig.~\ref{fig3}. Lines  correspond  to  numerical
solution  of the  exact  equations and  dots  correspond to  numerical
simulation on random graphs.}
\end{figure}
\end{center}

Figure \ref{fig5} shows the  behaviour of $U_{e}$ corresponding to the
same  four cases as  in Fig.~\ref{fig3}.   Below $\sigma_c$  the three
roots  of  Eq.~(\ref{past})   generate  a  lace  function.   Numerical
simulations  follow continuously one  of the  roots until  $H_2$ where
they jump to the lower exchange  energy branch.  Note that there is an
intermediate field $H_3$ (crossing point of the lace) where two of the
roots correspond to the same value of $U_{e}$.

\begin{center}
\begin{figure}[th]
\includegraphics[width=8.5cm,clip]{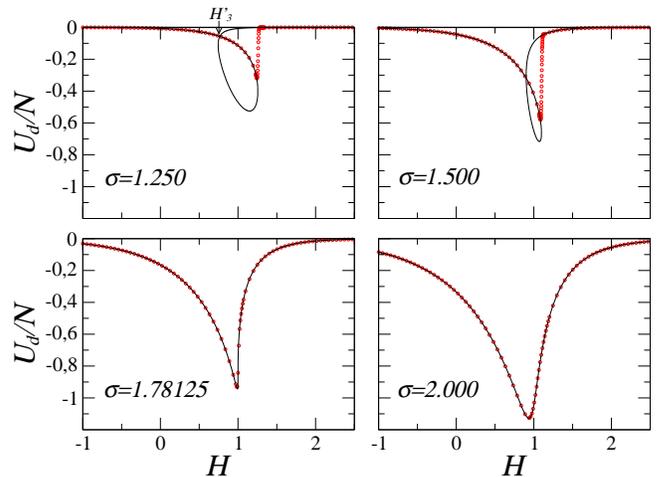}
\caption{\label{fig6}  Disorder-coupling energy  corresponding  to the
same  cases  as   in  Fig.~\ref{fig3}. Lines  correspond  to  numerical
solution  of the  exact  equations and  dots  correspond to  numerical
simulation on random graphs.}
\end{figure}
\end{center}

Figure \ref{fig6}  shows the behaviour  of $U_{d}$ \footnote{  We note
that this  energy term,  can be explicitely  written as  an analytical
function  of $P^\ast$,  but we  have omitted  it for  simplicity.}.  A
similar lace  shape is  obtained.  We note,  that the  crossing points
$H_3'$  (for $\sigma  < \sigma_c$)  are  different from  $H_3$ in  the
$U_{e}$  curve. As  $\sigma$  tends to  $\sigma_c$  the fields  $H_1$,
$H_2$,  $H_3$, and  $H_3'$  all  tend to  the  critical value  $H_c=1$
\cite{Dhar1997}.

\begin{center}
\begin{figure}[th]
\includegraphics[width=8.5cm,clip]{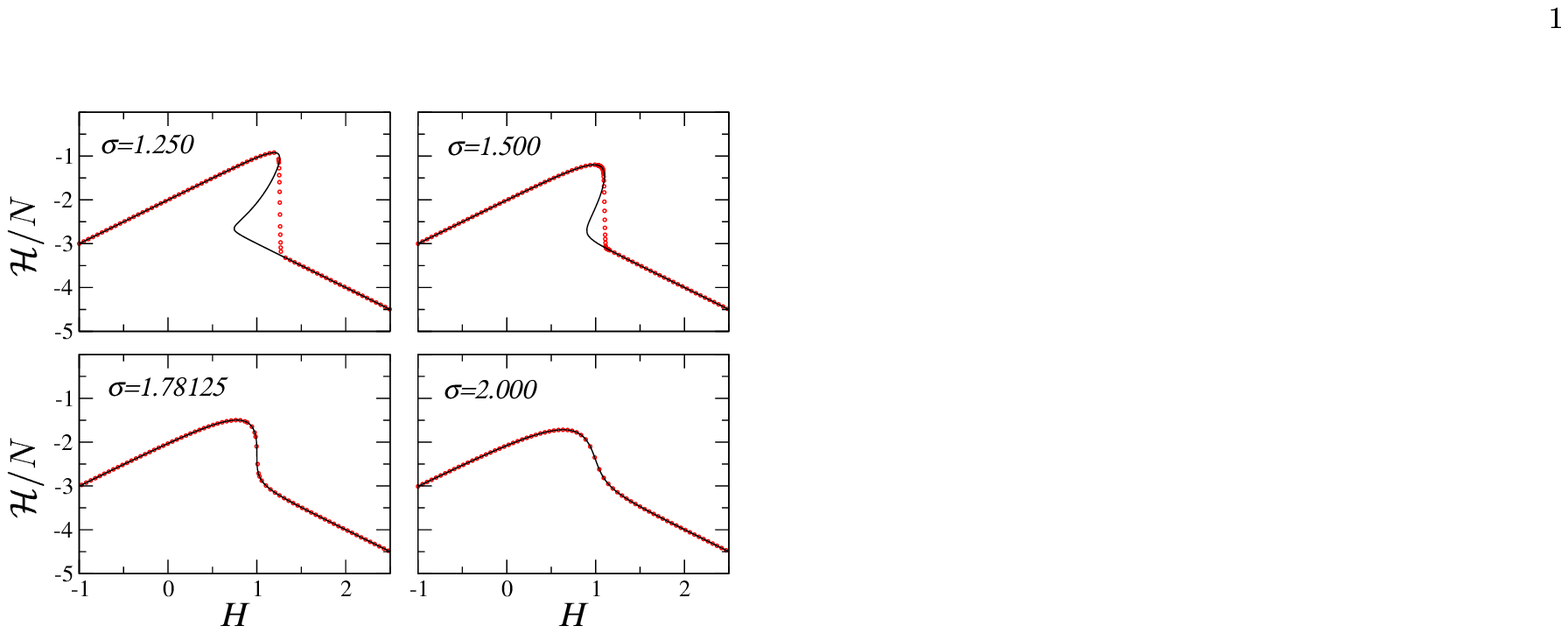}
\caption{\label{fig7}  Behaviour  of the  Hamiltonian  $\cal  H$ as  a
function  of the  external field  $H$ corresponding  to the  same four
cases  as  the previous  figures.   Numerical  solution  of the  exact
equations  are  shown  with   a  continuous  line,  whereas  numerical
simulations of random graphs are shown with a dotted line.}
\end{figure}
\end{center}

In  Fig.~\ref{fig7} we  show the  behaviour of  the  total Hamiltonian
(magnetic enthalpy) $\cal H$, corresponding  to the same four cases as
before.  The plots  show that, in the trivalued  region, the numerical
simulations choose the branch with  maximum $\cal H$, which is clearly
a non-equilibrium path.

\section{Energy dissipation}
\label{dissip}

\begin{center}
\begin{figure}[th]
\includegraphics[width=8.0cm,clip]{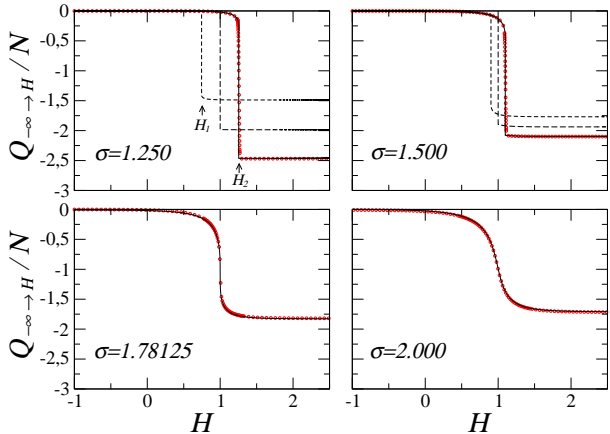}
\caption{\label{fig8}Energy  dissipated   along  the  metastable  path
$-\infty  \rightarrow H$, computed  from Eq.\  (\ref{eq:e-diss}).  For
$\sigma  <  \sigma_c$ we  compare  three  possible trajectories  which
differ  in the  field at  which the  transition occurs  (see  text for
details).   The dotted  line is the  result of  our numerical
simulations on random graphs. }
\end{figure}
\end{center}

The  fact of  having obtained  separately  $m$, $U_e$,  and $U_d$,  as
functions of $H$, allows an exact computation of the energy dissipated
by  the system  $Q$. It  can be  calculated by  integration  along the
non-equilibrium path \cite{Ortin1998}:
\begin{equation}
Q = \Delta U  - \int H dM,
\label{eq:dif-e-diss}
\end{equation}
where $\Delta U = \Delta U_e + \Delta U_d$ (internal energy difference
between the  initial and final  states). Starting from $H  = -\infty$,
the path integral over the trajectory computed in the previous section
can be written in the form:
\begin{equation}
\frac{Q(-\infty  \rightarrow H')}{N}  = \frac{U(H')}{N} + \frac{1}{2}zJ
-\int_{-\infty}^{H'}  H \  dm .
\label{eq:e-diss}
\end{equation}
Figure \ref{fig8}  shows the result obtained from  this expression for
$z=4$ and  different values  of $\sigma$.  For  those cases  for which
$\sigma < \sigma_c$ the energy  dissipated has been computed for three
different trajectories (indicated by  dashed or continuous lines), all
of  them  compatible  with  the theoretical  (numerical)  solution  of
$m(H)$,  $U_e(H)$,  and  $U_d(H)$:  the  first one  assumes  that  the
transition to the upper branch of $m(H)$ takes place at $H = H_1$, the
second one at an intermediate value  of the field $H = (H_1 + H_2)/2$,
and the third one at $H = H_2$.  Of these trajectories, we verify that
the one jumping  at $H_2$ (the trajectory chosen by  the system in the
numerical simulations)  is the one of maximum  energy dissipation ($Q$
most negative).

\begin{center}
\begin{figure}[th]
\includegraphics[width=8.5cm,clip]{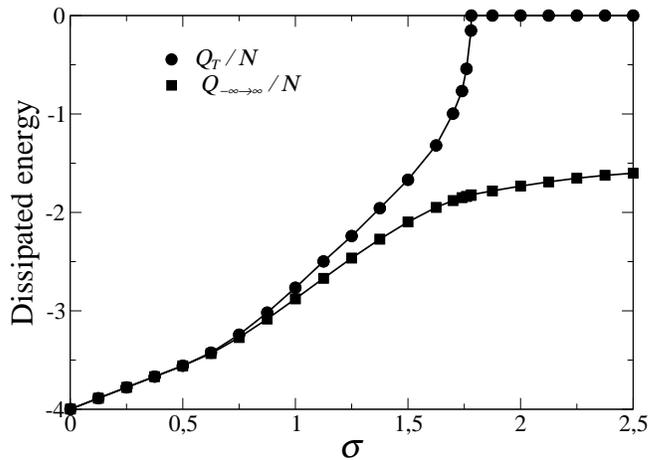}
\caption{\label{fig9}Comparison  of  the   energy  dissipated  at  the
transition  jump  $Q_T$ ({\Large  $\bullet$})  with  the total  energy
dissipated  along  the  full  path  $Q_{-\infty  \rightarrow  \infty}$
($\blacksquare$).}
\end{figure}
\end{center}

In Fig.~\ref{fig9}  we compare the energy  dissipation associated with
the magnetization  transition jump at  $H_2$ ($Q_T=\Delta {\cal  H} $)
with  the total  energy  dissipated along  the  full path  $Q_{-\infty
\rightarrow \infty}$.  One can  see that for  $\sigma <  \sigma_c$ the
dissipation at the transition $Q_T$  represents a large fraction of the
total dissipation.

\section{Summary  and conclusions}
\label{Conclusions}
We have studied  the RFIM on a Bethe lattice  with a 1-spin-flip local
relaxation  metastable  dynamics,  following  the method  proposed  in
Refs.~\onlinecite{Dhar1997,Shukla2001}.    We    have   extended   the
calculations  and   computed  the   different  energy  terms   in  the
Hamiltonian which  account for the spin-spin  exchange energy ($U_e$),
the spin-random field coupling term ($U_d$), and the energy associated
with the  external driving  field ($-HM$). The  analysis of  the Bethe
lattice  with coordination  numbers $z>3$  allows to  understand (with
analytic  equations)  the role  played  by  each  energy term  in  the
disorder induced phase transition that separates the phase with smooth
hysteresys loop from the phase with discontinuous hysteresis loop. The
availability  of the  separate energy  terms allows  the study  of the
energy  dissipation as  a function  of  the external  field along  the
hysteresis loop.
 
\section{Acknowledgements}
We acknowledge  fruitful comments from Prabodh  Shukla.  This research
has received financial  support through projects MAT2004-01291 (CICyT,
Spain),   BQU2003-05042-C02-02,  BFM2003-07749-C05-04   (MEC,  Spain),
SGR-2000-00433, SGR-2001-00066 (DURSI,  Generalitat de Catalunya).  We
aknowledge  a supercomputing  project from  CESCA  (Catalunya). X.Illa
acknowledges financial support from DGI-MEC.

\end{document}